\documentclass[aps,prl,showpacs,amsmath,amssymb,twocolumn,superscriptaddress]{revtex4-1}

\usepackage{graphicx}
\usepackage{color}
\usepackage{amsmath}

\begin{document}

\title{Non-equilibrium and proximity effects in superconductor-normal metal junctions}

\author{V. J. Kauppila}
\email[]{ville.kauppila@aalto.fi}
\author{H. Q. Nguyen}
\author{T.~T.~Heikkil\"a}
\affiliation{O.V. Lounasmaa Laboratory, Aalto University, P.O. Box 15100, FI-00076 AALTO, Finland}

\newcommand{\tmpnote}[1]%
   {\begingroup{\it (FIXME: #1)}\endgroup}
   \newcommand{\comment}[1]%
       {\marginpar{\tiny C: #1}}

\date{\today}

\begin{abstract}
We study the consequences of non-equilibrium heating and inverse proximity effect in normal metal - insulator - superconductor - insulator - normal metal (NISIN) junctions with a simple quasi-one-dimensional model. We especially focus on observables and parameter regions that are of interest in the design of SINIS coolers with quasiparticle traps. We present numerical results calculated by solving the Usadel equation and also present analytical approximations in two limiting cases: a short junction with a non-negligible resistance in both ends and a long junction with a transparent contact at one end.
\end{abstract}

\pacs{}

\maketitle

\section{Introduction}

Electronic on-chip coolers offer a promising method for reaching below-100$\,\rm{mK}$ temperatures in variety of applications including final stage of cooling in low temperature physics experiments as well as extremely sensitive radiation detectors used in space applications. A promising way to realize this type of cooler is by using a superconductor (S) weakly coupled to the normal metal (N) which is to be cooled \cite{giazotto2006opportunities}.

The operational principle of these NIS coolers is based on the presence of the energy gap in the superconductor density of states. When the bias voltage applied across the junction is adjusted correctly, only the hot quasiparticles can tunnel to the superconductor, thus cooling the normal metal. A more comprehensive review of the theory of NIS junctions can be found from Refs. \cite{giazotto2006opportunities, 0034-4885-75-4-046501}.

The history of (SI)NIS coolers dates back to the nineties \cite{0034-4885-75-4-046501}. While the understanding of these systems has progressed, it has been understood that the main limitation to the cooling is often due to the non-equilibrium heating of the superconductor \cite{pekola2000trapping}. A popular solution to bypass this limitation is to use another normal metal, a ``quasiparticle trap'', in contact with the superconductor to allow thermalization of the hot non-equilibrium quasiparticles in the superconductor \cite{pekola2000trapping, irwin1995quasiparticle}. Later the same effect was achieved by using magnetic fields \cite{peltonen2011magnetic}, where the mechanism is essentially the same with normal metallic vortex cores acting as quasiparticle traps. Recently it has also been demonstrated that making the superconductor wide close to the contact reduces the non-equilibrium heating \cite{knowles2012probing}. However, this approach is not applicable in systems with wide junctions aiming at large cooling powers.

So far the non-ideal characteristics of NIS coolers has been analyzed with simplified thermal models, which assume the presence of a quasiequilibrium distribution inside the superconductor, and do not include for example the inverse proximity effect. In this paper, our aim is to provide a microscopic description of the role of non-equilibrium effects in NIS coolers in the presence of a quasiparticle trap. We base our description on a quasi-one dimensional model and take into account the inverse proximity effect which also affects the cooling process in setups with well-coupled traps. Effects we do not take into account since they have been discussed elsewhere include the effect of the environment on the density of states \cite{pekola2010environment} and coherent \cite{rajauria2008andreev} and incoherent \cite{laakso2012manifestly} Andreev effects.

\section{The Model}

\begin{figure}[!ht]
\centering
\includegraphics[width=.99\columnwidth]{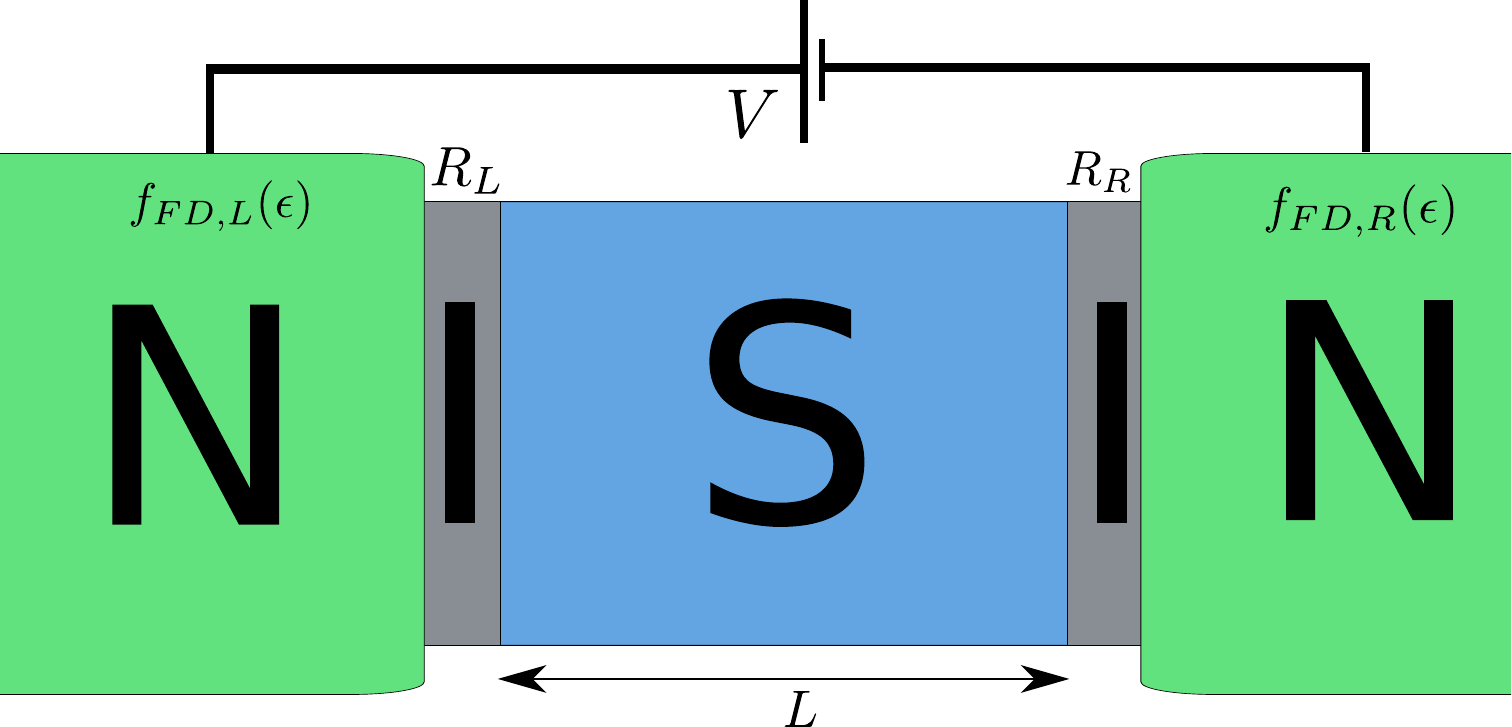}
\caption{\small Schematic picture of the quasi-one dimensional model. $f_{FD,L/R}$ are the Fermi-Dirac distribution functions in the left/right normal metals.}
\label{fig:NISN}
\end{figure}

We model the NIS cooler with a quasi-one dimensional model shown in Fig. \ref{fig:NISN}.
The model consists of a normal metal island which is to be cooled (on the left) in equilibrium with some temperature $T_L$, the normal metallic quasiparticle trap (on the right) in equilibrium with some temperature $T_R$ and the superconducting layer of length $L$ between the two. The superconducting layer is coupled to the two normal metals by insulating barriers with resistances $R_L$ and $R_R$, similar to the geometry in \cite{nquyenAPL}. The system is then biased with some voltage $V$. This model can also be used to study the NIS junction in the absence of the quasiparticle trap. In that case the length of the superconductor, $L$, must be set equal to the energy relaxation length of the superconductor.

The effects we take into account in our calculations are 1) the inverse proximity effect from the normal metals to the superconductor, 2) the non-equilibrium heating of the superconducting wire and 3) the electron-phonon interaction in the normal metal island, which also adds two parameters to our theory: the electron-phonon coupling strength $\Sigma$ multiplied by the volume $\Sigma$ of the island and the phonon temperature $T_{ph}$.

To make the calculations tractable, we consider two type of coolers separately. The first type is one with a sufficiently long wire (in practice $L > 3-4 \xi$, where $\xi$ is the coherence length of the superconductor) and a good contact between the quasiparticle trap and the superconductor, i.e. $R_R = 0$. The second case considers a short superconducting wire ($L < \xi$) and arbitrary resistances on the two interfaces.

Below, we switch to units where lengths are in the units of the superconductor coherence length, $\xi$, energies are in the units of the superconducting gap $\Delta_0$ at zero temperature and the resistances are in the units of the resistance of one coherence length long superconductor in the normal state, $R_\xi$. We also set $k_B = \hbar = e = 1$ so that the temperatures and voltages are also expressed in the units of energy. We retain SI units in the most important results.

The starting point of our analysis is the Usadel equation \cite{usadel1970generalized} which gives the quasiclassical Keldysh Greens functions of a superconductor in the dirty limit. The matrix Usadel equation can be further divided to two separate equations \cite{belzig1999quasiclassical}: the kinetic part which gives the distribution functions in the superconductor and the spectral part which gives the spectral coefficients needed to solve the first one. The kinetic Usadel equation for the distribution functions of the superconductor reads
\begin{equation}
\begin{aligned}
\partial_x \left( \mathcal{D}_L(x) \partial_x f_L (x) \right) & = 0 \\
\partial_x \left( \mathcal{D}_T \partial_x f_T (x) \right) & = f_T (x) 2 \mathcal{R} ,
\end{aligned}
\label{keldyshusadel}
\end{equation}
where we have divided the full distribution function of the superconductor to its symmetric and antisymmetric (or transversal and longitudal) parts in energy, $f_T$ and $f_L$, respectively. The full distribution function can be obtained from these by $f = \frac{1}{2}\left(1- f_L -f_T\right)$. The coordinate and energy dependent spectral coefficients appearing in Eq. \eqref{keldyshusadel} are the energy dependent spectral energy and charge diffusion coefficients, $\mathcal{D}_L = \cos^2(\operatorname{Im}\theta)$, $\mathcal{D}_T = \cosh^2 ( \operatorname{Re}\theta)$, and charge recombination coefficient, $\mathcal{R} = -\operatorname{Im}\sinh(\theta)$. These can be parametrized in terms of the pairing angle, $\theta(\epsilon, x)$, which can be solved from the spectral Usadel equation
\begin{equation}
\frac{\partial^2 \theta}{\partial x^2} = 2 i \Delta \cosh(\theta) - 2 i \epsilon \sinh(\theta) .
\label{thetaeq}
\end{equation}
Generally, Eq. \eqref{thetaeq} should also contain, and be accompanied with another equation containing phase difference, but for simplicity, we assume that the supercurrent flowing through the superconductor is negligible so that phase gradient can be discarded \cite{anthore2003density}. The presence of the phase gradient would also add extra terms to the kinetic equation. The three Usadel equations, two for the distribution functions and one for the pairing angle, are accompanied by six boundary conditions, which read \cite{kupriyanov}
\begin{equation}
\begin{aligned}
R_L \mathcal{D}_i(0) (\partial_x f_i(0))& = \left[ f_i(0) -f_i^0(\epsilon,\mu_L,T_L)\right] N_s(0) \\
R_R \mathcal{D}_i(L) (\partial_x f_i(L))& = \left[ f_i^0(\epsilon,\mu_R,T_R)-f_i(L) \right] N_s(L) \\
R_j \partial_x \theta \vert_{x = 0,L} & = \sinh(\theta(x=0,L)) ,
\end{aligned} ,
\label{kuprluk}
\end{equation}
where $N_S$ is the reduced density of states of the superconductor and $i \in \{L,T\}$. The reservoir distribution functions, $f_i^0$ are given by the symmetric and antisymmetric parts of the Fermi-Dirac distributions. It should be noted that due to the inverse proximity effect, the reduced density of states in general differs from the bulk value calculated from the BCS theory, i.e.
\begin{equation}
N_{S0} = \operatorname{Re}\frac{\epsilon}{\sqrt{\epsilon^2 - \Delta^2}} .
\end{equation}

After solving the three equations \eqref{keldyshusadel} and \eqref{thetaeq} either numerically or analytically after some approximations, we can calculate the energy and charge current densities from the normal metal island by
\begin{equation}
\begin{aligned}
j_E = &\int_{-\infty}^\infty d \epsilon \epsilon j_L (x=0) \\
j_c = &\int_{-\infty}^\infty d \epsilon j_T (x=0) ,
\end{aligned}
\end{equation}
where $j_i = \mathcal{D}_i \partial_x f_i$ is the spectral current. The heat current from the island is then given by
\begin{equation}
j_Q = j_E - \mu_L j_c .
\end{equation}

In what follows, we disregard the self-consistency of the superconducting gap. The temperature dependence of the gap can be safely neglected since we are considering only temperatures $T_L,T_R \ll \Delta$. Also it has been shown \cite{voutilainen2005nonequilibrium} that when $R_R \ll R_L$, the non-equilibrium effects are not expected to change the gap considerably. This is also the limit at which the cooling is expected to work best and we may safely limit the calculations below to this limit. Furthermore, due to the proximity effect the gap close to the quasiparticle trap would be suppressed, which amounts to only renormalizing the length of the superconductor, leaving the qualitative results unchanged.

\section{Density of states}

As stated above, the most important effect from the proximity of the normal metal is to induce subgap states in the superconductor. To this end, we give an analytic expression for the proximity modified density of states in the two cases described above.

\subsection{Long wire, direct trap cooler}

In this case, we have two contributions to the density of states that can be separated. First is the modification due to the right (good) interface and the second is the modification due to the left (high resistance) interface. To calculate the former one, we solve the spectral Usadel equation \eqref{thetaeq} in the limit of a long wire and $R_L \gg 1$, $R_R = 0$. The result is
\begin{equation}
\theta(x) \approx \theta_S - 4 \operatorname{arctanh}\left[\exp\left(-(L-x) \sqrt{2\alpha}\right) \tanh(\theta_S/4)\right] ,
\label{semiinftheta}
\end{equation}
where $\theta_S = \rm{arctanh}\left(\Delta/\epsilon\right)$ is the pairing angle of a bulk superconductor and $\alpha = \sqrt{\Delta^2 - \epsilon^2}$. This solution describes a semi-infinite wire with a good contact to a normal metal reservoir at $x=L$. The density of states can now be calculated from $N_S = \operatorname{Re}\cosh\theta$. With the pairing angle \eqref{semiinftheta} this becomes
\begin{equation}
N_S = N_{S 0} +  \operatorname{Re} \frac{\epsilon(1+6\beta^2+\beta^4)-4 \Delta(\beta+\beta^3)}{\sqrt{\epsilon^2-\Delta^2}(1-\beta^2)^2} ,
\end{equation}
where $\beta = \exp(-L\sqrt{2\alpha})\tanh(\theta_S/4)$. By expanding the second term for small $\beta$, we get outside the region $| \Delta - \epsilon | < 1/(2 \Delta L^4)$
\begin{equation}
N_S = N_{S,0} - \operatorname{Re}\frac{4 \Delta e^{-L \sqrt{2 \alpha}} \tanh(\theta_S/4)}{\sqrt{\epsilon^2-\Delta^2}} .
\label{nslapprox}
\end{equation}
At zero energy, this becomes $N_S(\epsilon = 0) \approx 1.7 e^{-L \sqrt{2}}$. 

The latter contribution, i.e., the contribution due to the finite (but large) resistance in the left interface can be calculated by expanding $\theta$ around its bulk value and then solving the spectral Usadel equation. The density of states from this becomes
\begin{equation}
N_S(x=0) \approx N_{S0} + \operatorname{Re}\frac{\Delta^2}{\sqrt{2} R_L (\Delta^2-\epsilon^2)^{5/4}} .
\end{equation}

\begin{figure}[!ht]
\centering
\includegraphics[width=.99\columnwidth]{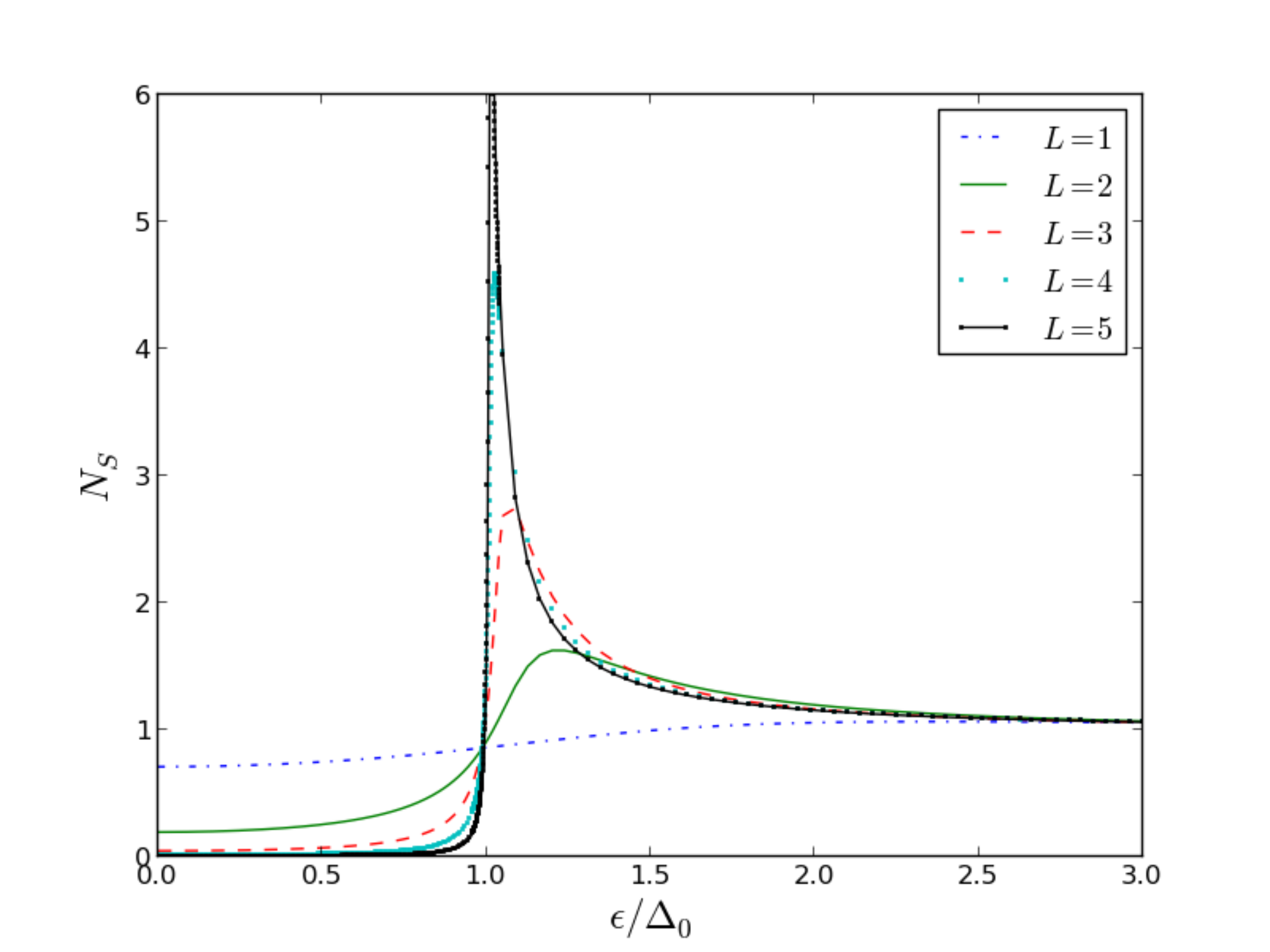}
\caption{\small (Color online) Density of states at the left end of the wire as a function of energy with $R_L = 10000$ and $R_R = 0$.}
\label{fig:dos}
\end{figure}

When the deviation of the pairing angle from the bulk value is small, these two contributions can be added to give the total density of states in this geometry. The rounding of the density of states for different lengths of the wire using an exact numerical solution for some parameter values is shown in Fig. \ref{fig:dos}. 

\subsection{Short wire, arbitrary resistances}

When $L \ll 1$, we can calculate the pairing angle as follows. We approximate the derivative in Eq. \eqref{thetaeq} with a difference, $\partial_x^2 \theta \approx (\partial_x \theta(x=L) - \partial_x \theta(x=0))/L$. Then we use the boundary conditions and assume that $\theta$ does not depend on $x$. The solution becomes
\begin{equation}
\theta = \operatorname{arctanh}\left(\frac{\Delta}{\epsilon + \frac{i}{2 L \tilde{R}}}\right) \equiv \operatorname{arctanh}\left(\frac{\Delta}{\epsilon + i \gamma}\right)  ,
\end{equation}
where $\tilde{R} = R_L R_R / (R_L + R_R)$ is the parallel resistance of the two barriers and $\gamma \equiv \Delta \xi R_\xi / (2 L \tilde{R})$ is the Dynes factor. The density of states from this becomes of the Dynes form
\begin{equation}
N_S = \operatorname{Re}\frac{(\epsilon + i \gamma)^2}{\sqrt{(\epsilon + i \gamma)^2 - \Delta^2}} .
\end{equation}
It is worth noting that in this limit, the Dynes parameter is independent of energy.

\section{Heat current}

The heat current from the normal metal in an ideal NIS junction (i.e., without taking into account the non-equilibrium heating or the proximity effect), assuming the superconductor and the normal metal are both in equilibrium with some temperatures $T_L$, $T_R$ and distribution functions given by Fermi-Dirac distributions, $f_{FD}$, reads \cite{anghel2001noise}
\begin{widetext}
\begin{equation}
\begin{aligned}
j_Q^{NIS} & = \frac{1}{R_L} \int d\epsilon (\epsilon - V) N_{S0} \left(f_{FD}(\epsilon, V, T_L) - f_{FD}(\epsilon, 0, T_R)\right) \\
& \approx \frac{\Delta^2}{R_L}\sqrt{\frac{\pi T_L^3}{2 \Delta^3}} \left[ \frac{1}{2} g_{3/2}\left(\frac{\Delta-V}{T_L}\right)+\frac{\Delta-V}{T_L} g_{1/2}\left(\frac{\Delta-V}{T_L}\right) \right] - \frac{\Delta^2}{R_L} \sqrt{\frac{2 T_R \pi}{\Delta}} e^{-\Delta/T_R} ,
\end{aligned}
\label{jqnis}
\end{equation}
\end{widetext}
where the approximation holds when $V + T_L < \Delta$. We now proceed to considering the two additional modifications to this -- non-equilibrium and proximity effect -- separately as corrections to this formula in the two geometries descibed above.

\subsection{Inverse proximity effect}

For a moment we forget the non-equilibrium effects and write the heat current as in the ideal NIS case, but with the proximity modified density of states, i.e.
\begin{equation}
\begin{aligned}
j_Q^{PE} = & \frac{1}{R} \int d\epsilon (\epsilon - \mu_L) N_{S} \\
& \times \left(f_{FD}(\epsilon, \mu_L, T_L) - f_{FD}(\epsilon, \mu_R, T_R)\right) ,
\end{aligned}
\end{equation}
where $R = R_L + R_R$ and $\mu_L-\mu_R = V$. For voltages $V + T_L \ll \Delta$, we can expand the density of states at $\epsilon \approx 0$ and calculate the integral. Formally the heat current density can be written as
\begin{equation}
j_Q^{PE} = j_Q^{NIS} + \delta j_Q^{PE} ,
\end{equation}
where the first term is the ideal NIS current given above and the second term is a correction to this.

In the first type of a geometry with a direct trap and a long wire using the zero-energy expansion the correction becomes
\begin{equation}
\begin{aligned}
\delta j_Q^{PE} \approx & \frac{2 \Delta^2}{e^2 R_L} \tan\left(\frac{\pi}{8}\right) e^{-\frac{L}{\xi} \sqrt{2}} \\
& \times \left[\frac{k_B^2\pi^2}{3\Delta^2} \left(T_L^2 - T_R^2 \right) -  \left(\frac{eV}{\Delta}\right)^2\right] .
\end{aligned}
\label{jpeappr}
\end{equation}

In the second type of a geometry, i.e. $L \ll 1$, we get after identical considerations, in the relevant limit $R_L \gg R_R$
\begin{equation}
\delta j_Q^{PE} \approx \frac{\Delta \gamma}{e^2 R} \left[ \frac{\pi^2 k_B^2}{6 \Delta^2} (T_L^2 - T_R^2) - \frac{1}{2} \left(\frac{eV}{\Delta}\right)^2\right] .
\end{equation}

Next we proceed to consider the effect of the non-equilibrium heating.

\subsection{Non-equilibrium heating}

Now we need the non-equilibrium expressions for the heat current that can be solved from Eq. \eqref{keldyshusadel}. First we calculate them in the first type of a geometry with a long wire and $R_R = 0$. We approximate the spectral coefficients by their bulk values and solve Eqs. \eqref{keldyshusadel}. For the longitudal spectral current density at the interface of the island and the superconductor we get
\begin{equation}
R_L j_L = \frac{N_{S0} \delta f_L(\epsilon)}{1 + \frac{L}{R_L} \mathcal{D}_L N_S}
\label{jlneq}
\end{equation}
and for the transversal current density
\begin{equation}
R_L j_T = - \frac{N_{S0} f_T(\epsilon, V)}{1 + \frac{N_S \tanh(\sqrt{\mathcal{R}/\mathcal{D}_T} L)}{\mathcal{D}_T R_L}} .
\end{equation}
Analysis of these shows that the modification to the transversal current density vanishes when $\epsilon = \Delta$ and is proportional to $\sim L/R_L$ otherwise and can thus be neglected since $L/R_L \ll 1$ for a typical setup. The correction to $j_L$ in expression \eqref{jlneq}, however, while also proportional to $\sim L / R_L$, diverges when $\epsilon \approx \Delta$ and thus needs to be taken into account. To calculate the correction, we expand the energy mode of the heat current, i.e. integral of $\epsilon$ times Eq. \eqref{jlneq}, in the limit of small $L/R_L$. The result becomes in the first order
\begin{equation}
R_L j_E \approx R_L j_E^{NIS} + R_L \delta j_E^{NEQ} ,
\end{equation}
where
\begin{equation}
R_L \delta j_E^{NEQ} = - \frac{L R_\xi}{R_L \xi} \int_{\Delta+\nu}^\infty \frac{\epsilon^3}{\epsilon^2 - \Delta^2} \Delta f_L(\epsilon) .
\end{equation}
Here we have also introduced a cutoff energy, $\nu$, to cut off the logarithmic divergence of the integral at $\epsilon \rightarrow \Delta$. In the limit $V + 2T_{L/R} < \Delta$ we can approximate the distribution functions with exponential functions and get as a result
\begin{equation}
\delta j_E^{NEQ} \approx -\frac{\Delta^2}{e^2}\frac{L R_\xi}{2 R_L^2 \xi} e^{-\frac{\Delta-eV}{k_BT_L}} \log\left(\frac{k_B T_L}{\nu e^{\gamma}} \right) ,
\label{jneqappr}
\end{equation} 
where $\gamma \approx 1.78$ is the Euler's constant. The cutoff is chosen from the condition $(L/R_L)N_S \approx 1$, which gives $\nu \sim \Delta L^2 R_\xi^2 /(2 R_L^2 \xi^2)$.

As for the second type of geometry, i.e., $L \ll 1$ and $R_R > 0$, the same expression \eqref{jneqappr} holds, provided that $R_L \gg R_R$.

\subsection{Total heat current}

In the limit where the two corrections to the ideal NIS current are both small, we can get the total current from the NIS current by adding the two corrections together, i.e.
\begin{equation}
j_Q = j_Q^{NIS} + \delta j_Q^{PE} + \delta j_Q^{NEQ} .
\label{jqtot}
\end{equation}

\begin{figure}[!ht]
\centering
\includegraphics[width=.99\columnwidth]{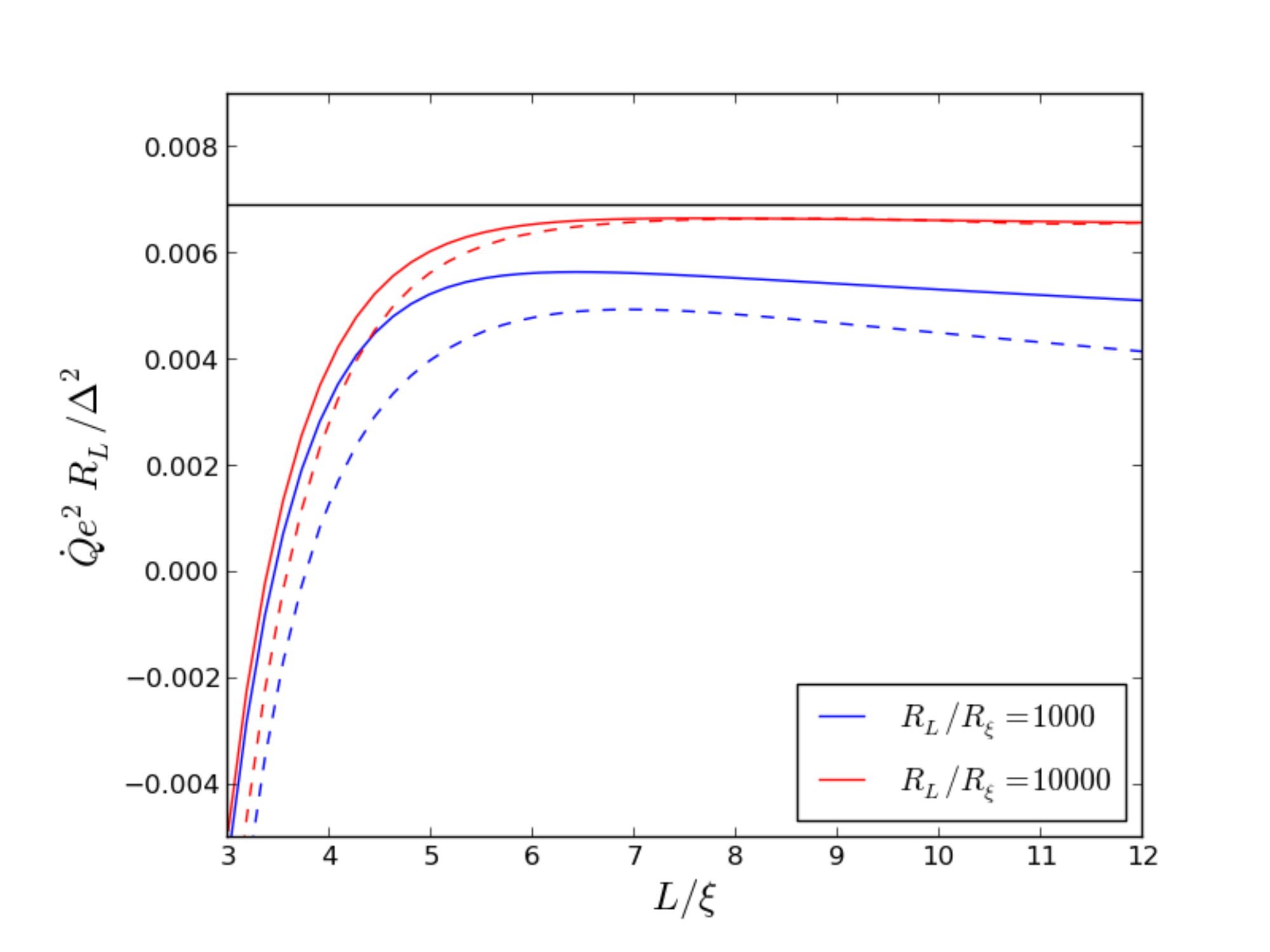}
\caption{\small (Color online) Heat current as a function of the length of the superconductor with $e V = 0.7 \Delta_0$ and $k_B T_L = k_B T_R = 0.1 \Delta_0$. The dashed line is the exact numerical result and the solid line is approximation \eqref{jqtot}. The black solid line is the ideal NIS current. The upper lines closer to the ideal NIS current (red) are for $R_L/R_\xi = 10000$ and the lower lines (blue) are for $R_L/R_\xi = 1000$.}
\label{fig:hcvsL}
\end{figure}

We first consider the total heat current for the setup with a long wire and a good contact between the trap and the superconductor. Since the proximity effect inhibits the cooling in short superconductors and non-equilibrium effect inhibits it in long superconductors, there exists a local maximum of the cooling as a function of the length of the superconductor. This behavior is shown in Fig. \ref{fig:hcvsL}. Along with this analytic approximation we have plotted the heat current calculated by numerically solving the Usadel equation. The main deviations from the numerical data are due to the breakdown of the semi-infinite approximation at small lengths of the superconductor.

\begin{figure}[!ht]
\centering
\includegraphics[width=.99\columnwidth]{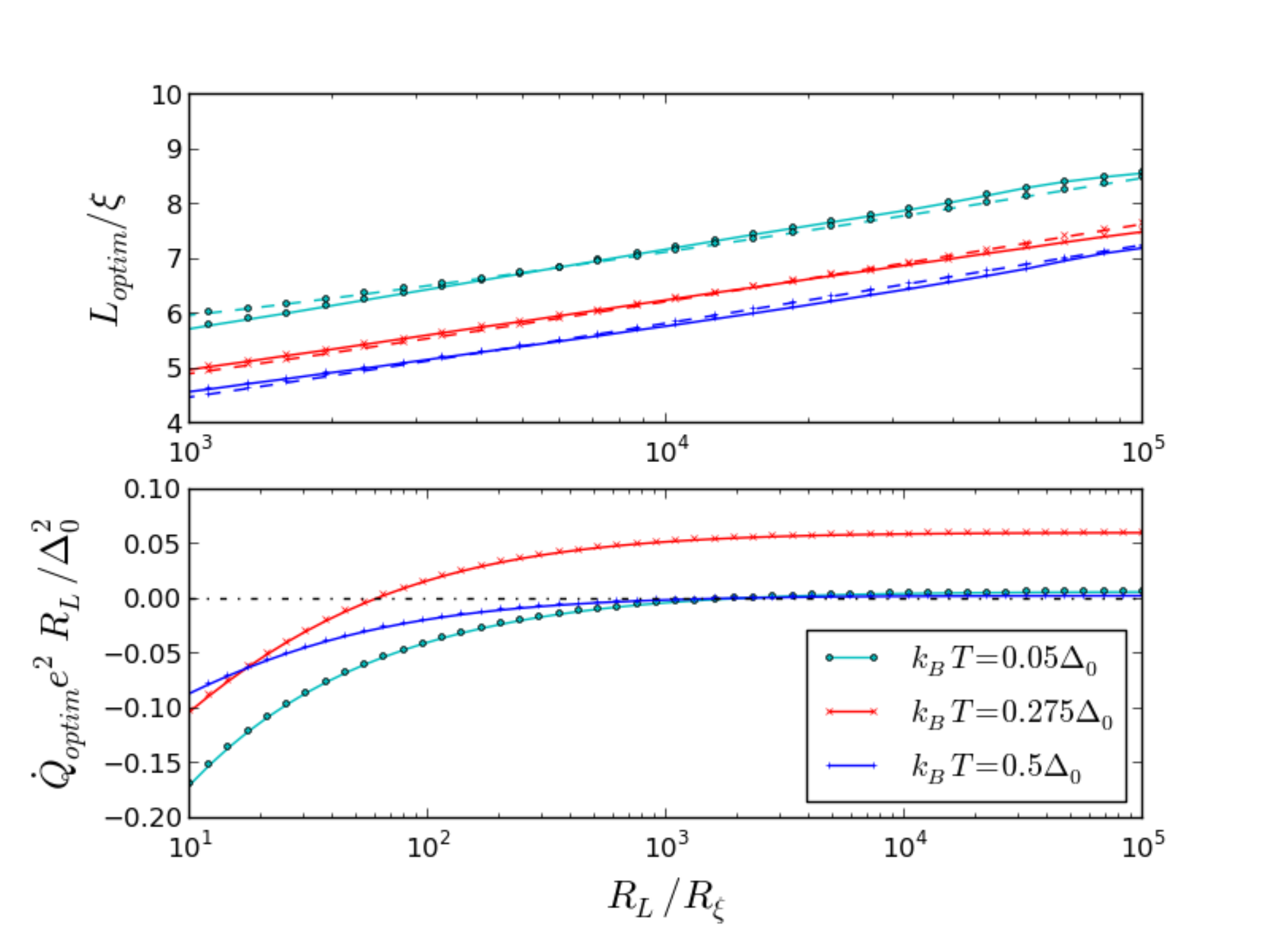}
\caption{\small (Color online) Numerically calculated optimal length (upper figure) and the heat current at that length (lower figure) as a function of the interface resistance at $e V = \Delta - 0.66 k_B T$ and $T_L = T_R \equiv T$. In the upper figure, we have also plotted the analytic approximation \eqref{Lopt} (dashed line) shifted by a constant to account for the discrepancy discussed in the text.}
\label{fig:Lopt}
\end{figure}

The optimal length of the wire is found from expression \eqref{jqtot}. It is
\begin{equation}
L_{opt} \approx \frac{\xi}{\sqrt{2}} \log \left[ \frac{4 \sqrt{2} \tan\left(\frac{\pi}{8}\right)\left(\frac{\pi^2}{3}\left(T_L^2 - T_R^2\right) - V^2 \right) R_L}{e^{-\frac{\Delta - V}{T_L}} \log \left(\frac{\delta e^\gamma}{T_L} \right) R_\xi}\right] .
\label{Lopt}
\end{equation}
This expression is expected to work when the assumptions made in the calculation of the heat current corrections above hold. The optimal length and the heat current at that length as calculated from the full numerics are shown in Fig. \ref{fig:Lopt}. For large $R_L$, the behavior corresponds roughly to Eq. \eqref{Lopt}, albeit there is a difference of $L \sim 1$ between the two, which results from our approximation to use the bulk values for the spectral quantities $\mathcal{D}_{L/T}$ and $\mathcal{R}$.

\subsection{Temperature of the island}

Next we consider the minimum temperature the normal metal can be cooled into. This temperature minimum can be found by solving for $T_L$ the heat balance equation
\begin{equation}
j_Q = 0 .
\end{equation}
For an ideal NIS junction, the limit is found to be $T_L^{min} \approx 1.42 \Delta \left(2 \pi T_R / \Delta \right)^{1/3} \exp (-2 \Delta / 3 T_R)$. In practice, however, the non-idealities limit the minimum temperature. We now consider these limitations.

\begin{figure}[!ht]
\centering
\includegraphics[width=.99\columnwidth]{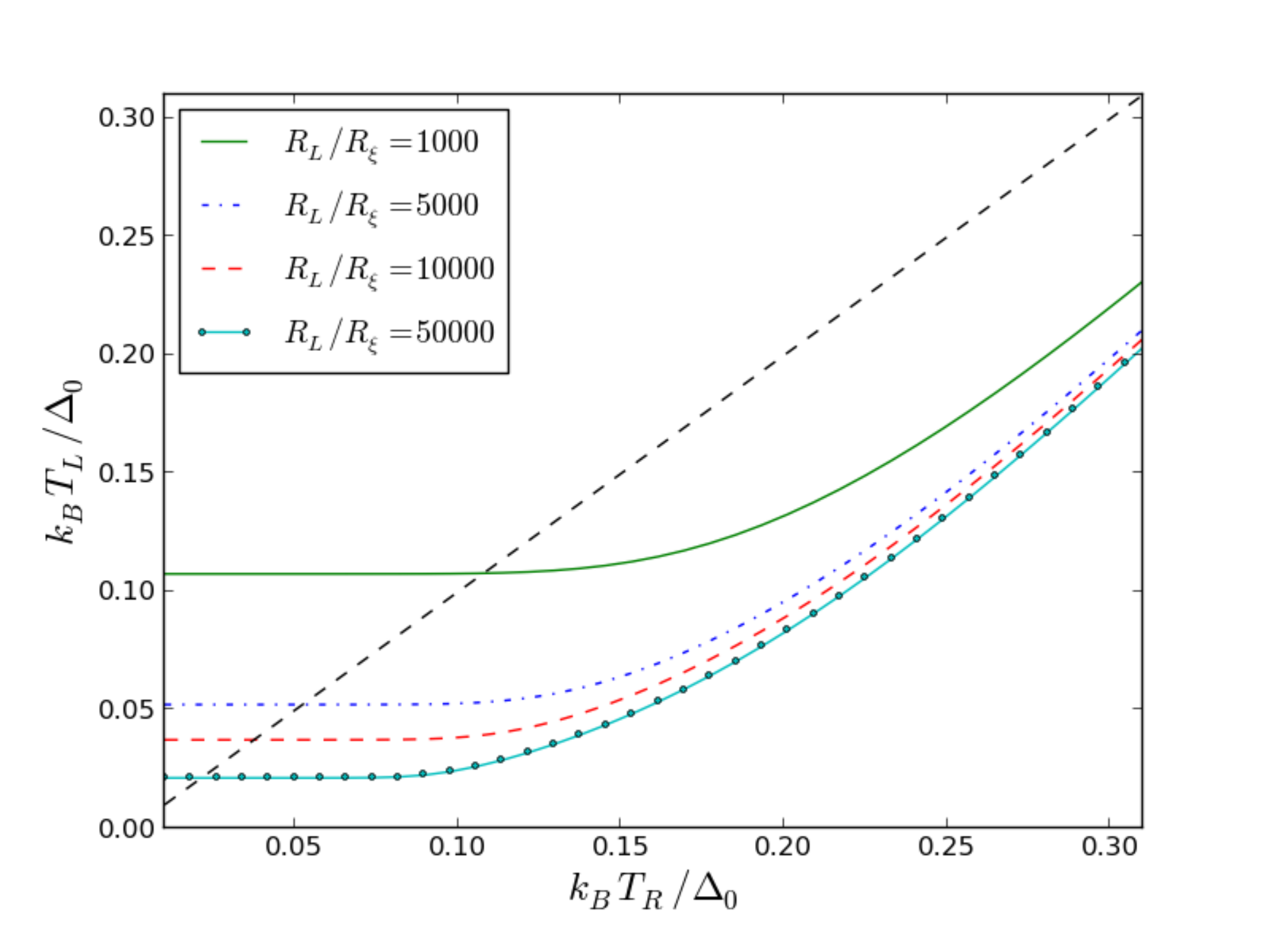}
\caption{\small (Color online) Minimum temperature of the normal metal island as a function of the temperature of the trap with $L=13 \xi$ and $e V \approx e V_{opt} = \Delta - 0.66 k_B T_L$. For large $T_R$, it is limited by the first term in Eq. \eqref{minTn1} whereas for low $T_R$ it saturates to the value given by the second term.}
\label{fig:minTn1}
\end{figure}

First we consider the long-wire setup. If the wire is long enough, the proximity effect does not matter and we can take into the heat current only terms $j_Q^{NIS}$ and $\delta j_Q^{NEQ}$. We also assume that the electron-phonon coupling is negligible and instead analyze it separately below. The minimum temperature becomes
\begin{equation}
\begin{aligned}
T_L \approx & \frac{\Delta}{k_B} \Bigg[A \sqrt{\frac{k_B T_R}{\Delta}} e^{-\frac{\Delta}{k_B T_R}} \\
& + \frac{L R_\xi}{R_L \xi} B \log\left(\frac{k_B T_L R_L^2 \xi^2}{\Delta L^2 R_\xi^2}\right)\Bigg]^{2/3},
\end{aligned}
\label{minTn1}
\end{equation}
where $A$ and $B$ are approximately constants at the optimal bias voltage when the ratio $(\Delta - eV)/ k_B T_L$ is fixed. $A$ comes from the NIS current expression and is approximately $A \approx 4.25$ at the optimal bias voltage. $B$ is given by the non-equilibrium correction to the current and its numerically found value is $B \approx 1.7$ at the optimal bias voltage. From expression \eqref{minTn1} we can see that at very low environment temperatures $T_R$, the minimum temperature is limited by $\sim\left(L \log(R_L/L) /R_L\right)^{2/3}$. Numerical simulations showing this behavior are shown in Fig. \ref{fig:minTn1}.

In the case of a very large resistance $R_L$, the limitations to the temperature are no longer due to the non-equilibrium heating but due to the electron-phonon coupling and thus this effect must be added to the heat current. The electron-acoustic phonon heat current is given by \cite{giazotto2006opportunities}
\begin{equation}
j_Q^{e-ph} = \sigma (T_L^5 - T_{ph}^5) ,
\end{equation}
where $\sigma \equiv \Sigma \Omega e^2 R_\xi \Delta^3 / k_B^5$ is the dimensionless electron-phonon coupling constant. In this case, as $T_R \rightarrow 0$, the minimum temperature is given by
\begin{equation}
T_L^{min} \approx \frac{\Delta}{k_B} \left(\frac{2 R_L^2 \Sigma^2 \Omega^2 e^4 \Delta^6}{\pi A^2 k_B^{10}} \right)^{1/3} \left(\frac{k_B T_{ph}}{\Delta}\right)^{10/3} .
\label{tephmin}
\end{equation}

\begin{figure}[!ht]
\centering
\includegraphics[width=.99\columnwidth]{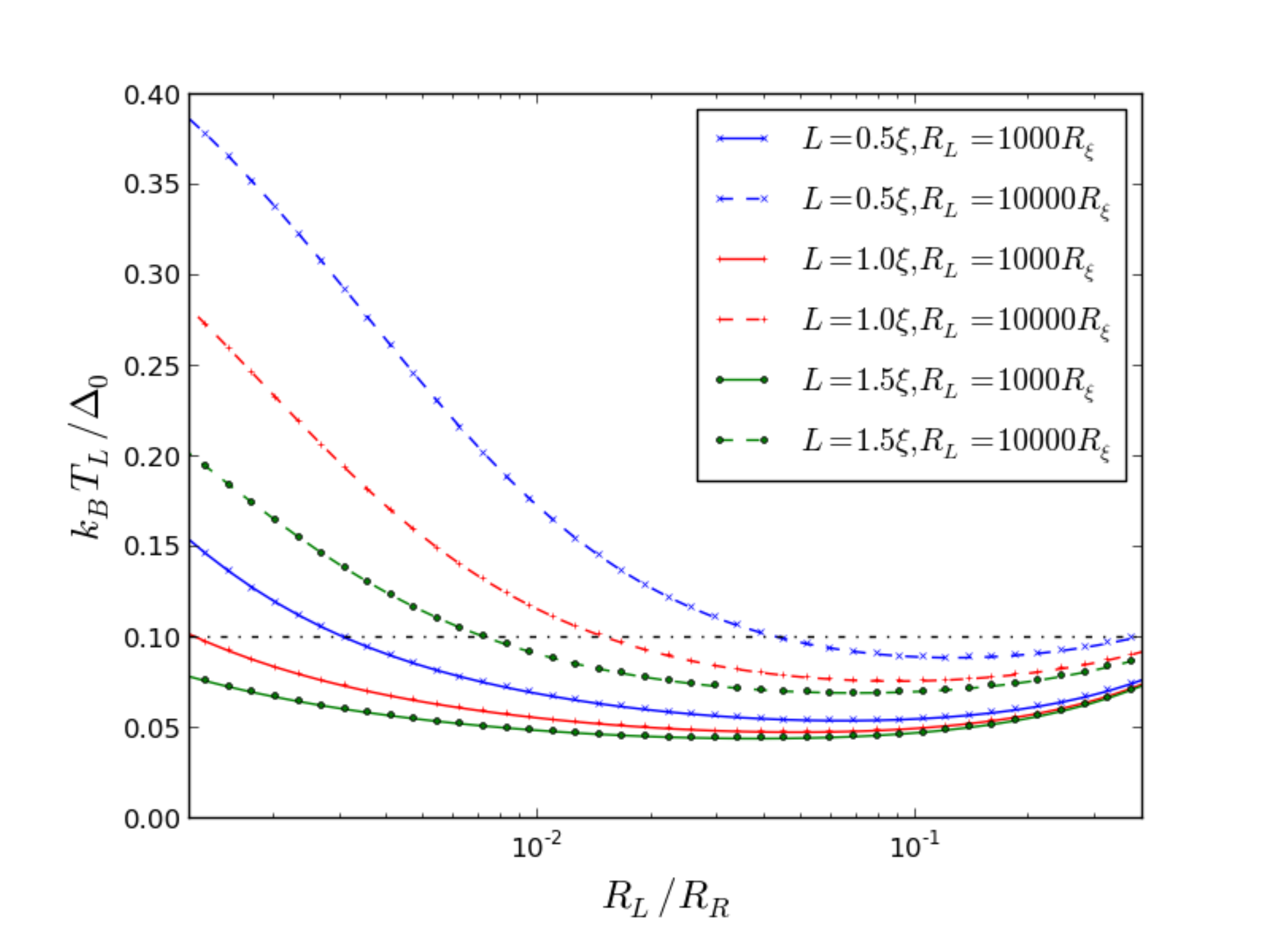}
\caption{\small (Color online) Equilibrium temperature of the normal island as function of $R_L/R_R$. Blue, red and green lines are $L/\xi = 0.5, 1, 2$ respectively. Solid lines are for $R_L / R_\xi = 10000$ and dashed lines are for $R_L / R_\xi = 1000$. The temperature of the trap is $k_B T_R / \Delta_0 = 0.1$.}
\label{fig:mintvsrshort}
\end{figure}

In the short junction setup we have different kind of a competition between the non-equilibrium and the inverse proximity effect. In the relevant limit $R_L \gg R_R$ the proximity effect always dominates the non-equilibrium term in the heat current. This can be seen by comparing the prefactors of the two and noticing that
\begin{equation}
R \delta j_Q^{PE} \propto \gamma \approx \frac{1}{2 L R_R} \gg \frac{L}{2 R_L} \propto R_L \delta j_Q^{NEQ} .
\end{equation}
However, if we make the right resistance larger, at some point the non-equilibrium heating starts to matter and then it limits the cooling. Thus there must be an optimum value of $R_L/R_R$ which maximizes the cooling. We show this behaviour in Fig. \ref{fig:mintvsrshort}.
These considerations are valid until we increase the total resistance of the junction enough so that the electron-phonon coupling starts to dominate. In this case, the minimum temperature is again given by Eq. \eqref{tephmin}.

\section{Charge current}
We repeat similar analysis as above for charge current. For an ideal NIS junction the charge current is given by
\begin{equation}
R I^{NIS} = \int d\epsilon N_{S0} (f_{FD}(\epsilon,V,T_L)-f_{FD}(\epsilon,0,T_R)) .
\end{equation}
For bias voltages $V \ll \Delta$ this becomes
\begin{equation}
R I^{NIS} \overset{\underset{eV \ll \Delta}{}}{\approx} \frac{\sqrt{2 \pi \Delta k_B T_L}}{e} e^{-\Delta/k_BT_L} \sinh\left(\frac{eV}{k_BT_L}\right) .
\end{equation}
For bias voltages $V \gg \Delta$ the NIS charge current reduces to the Ohms law plus a correction that vanishes at high voltages:
\begin{equation}
R I^{NIS} \overset{\underset{eV \gg \Delta}{}}{\approx} V - \frac{1}{2} \frac{\Delta^2}{e^2 V} .
\end{equation}

Using similar methods as above to calculate the heat current, we can calculate the corrections to NIS current due to non-equilibrium effects and the inverse proximity effects. We first calculate the long junction case. The non-equilibrium correction becomes (to the first order in $L/R_L$)
\begin{equation}
R_L \delta I^{NEQ} \approx
\begin{cases}
\frac{k_B T_L}{e} e^{-(\Delta-eV)/k_BT_L} \frac{LR_\xi }{R_L\xi} , & eV \ll \Delta \\
V \frac{LR_\xi}{R \xi} , & eV \gg \Delta \\
\end{cases} .
\end{equation}
For $V \gg \Delta$ the correction due to the proximity effect vanishes, but for $V \ll \Delta$ it becomes
\begin{equation}
R_L \delta I^{PE} \approx 4 \tan\left(\frac{\pi}{8}\right) e^{-L \sqrt{2} / \xi} V .
\end{equation}

For the short junction setup, we calculate only the proximity correction. For $V \ll \Delta$ it is given by
\begin{equation}
R_L \delta I^{PE} \approx \frac{2 \gamma}{e \Delta} \int d\epsilon f(\epsilon - \mu_L) = \frac{V \xi R_\xi^2}{L R_L R_R} .
\end{equation}

We note that unlike the heat current \eqref{jneqappr}, the non-equilibrium correction to the charge current at low bias voltages is proportional to the temperature, making it less important than in the case of the heat current. This can be traced back to the fact that the charge current is given by only the transversal part of the spectral current in which, as we noted above, the coefficient in front of the small parameter $L/R_L$ does not diverge at $\epsilon = \Delta$.

\section{Conclusions}

We have considered the heat and charge transport in NISIN junctions. We derived analytical estimates for the currents and temperatures of the normal metal island in two types of setups relevant for NIS cooler experiments: short junction with non-transparent contacts and long junction with a transparent contact between the superconductor and the quasiparticle trap. We discussed three different effects on the heat current: the inverse proximity effect to the superconductor, the non-equilibrium heating of the superconductor and the electron-phonon coupling.

%

\end{document}